\documentclass{llncs}

\usepackage[english]{babel}
\usepackage{alltt}
\usepackage{graphicx}
\usepackage{url}
\usepackage{subfigure}
\usepackage{listings}
\usepackage{color}
\usepackage{lscape}
\usepackage{xspace}
\usepackage{latexsym}
\usepackage{rotating} 
\usepackage{enumitem}

\usepackage{verbatim}
\usepackage{url}
\usepackage{multirow}

\usepackage{colortbl}
\usepackage[table]{xcolor}

\makeatletter
\renewcommand{\thetable}{\thesection.\@arabic\c@table}
\@addtoreset{table}{section}
\makeatother

\definecolor{MyDarkBlue}{rgb}{0,0.08,0.45}

\begin{document}

\title{Data Shapes and Data Transformations}
	\author{
	Michael Hausenblas\inst{1} 
	\and 
	Boris Villaz\'on-Terrazas\inst{2} 
	\and 
	Richard Cyganiak\inst{1}}
	\institute{DERI, NUI Galway, Ireland\\
	\email{firstname.lastname@deri.org}
	\and
	iSOCO, Madrid, Spain\\
	\email{bvillazon@isoco.com}}
\maketitle

\begin{abstract}
Nowadays, information management systems deal with data originating from different sources including relational databases, NoSQL data stores, and Web data formats, varying not only in terms of data formats, but also in the underlying data model.
Integrating data from heterogeneous data sources is a time-consuming and error-prone engineering task; part of this process requires that the data has to be transformed from its original form to other forms, repeating all along the life cycle. 
With this report we provide a principled overview on the fundamental data shapes \emph{tabular}, \emph{tree}, and \emph{graph} as well as transformations between them, in order to gain a better understanding for performing said transformations more efficiently and effectively. 
\end{abstract}

\section{Motivation}
\label{sec:mot}
These days, content and information management systems have to deal with data originating from an array of sources, such as relational databases, NoSQL data stores, and Web data formats. The data sources vary not only in terms of data formats, but first and foremost in the underlying data model, be it implicit---such as with JSON---or explicit, think: RDF. 

As recently put forward by Helland~\cite{Helland:ACMQ1}, data integration of heterogeneous data sources is a time-consuming, costly, and error-prone engineering task. Typically, the data has to be transformed from its original form to other forms, repeating all along the life cycle. For example, let us assume we want to publish data from an government agency such as spreadsheets containing statistical information into the Linked Open Data cloud\footnote{\url{http://lod-cloud.net}}.  One task of the Linked Data life cycles would then be to transform the original tabular spreadsheet data into (graph-shaped) RDF. Once we have the data transformed into RDF according to, say, the RDF Data Cube Vocabulary\footnote{\url{http://www.w3.org/TR/vocab-data-cube/}}, we want to visualize it in an appealing way, so we decide to use the Google Charts API\footnote{\url{https://developers.google.com/chart/}} requiring us to provide input as tabular data. So, again we have to transform a graph into a tabular, and then the application visualize the information in an appealing way.

Apparently, even in this toy example, we have to transform data from its original form to potentially many (intermediate) other forms. This was the motivation for us to compile this report, aiming to provide a principled overview of possible fundamental data shapes and transformations between them. In the following we will focus on the data transformation within the context of the ``Extract, Transform and Load'' process, a valuable process, growing in its use and scale. We use the term \emph{data shape} to refer to how the data is arranged and structured, closely related to the term data model\footnote{\url{http://en.wikipedia.org/wiki/Structured_data}}: we have identified three fundamental \emph{data shapes}: tabular, tree, and graph and respective transformations between them.

The remainder of the report is organized as follows: in Section~\ref{sec:fds} we introduce and motivate the fundamental data shapes, then in Section~\ref{sec:fct} we describe transformations between data shapes, and, finally, in Section~\ref{sec:rac} we discuss open issues and challenges concerning the transformations.

\section{Fundamental data shapes}
\label{sec:fds}
In the following we motivate and introduce the three fundamental data shapes \emph{tabular}, \emph{tree}, and \emph{graph}, derived from data structures\footnote{\url{http://en.wikibooks.org/wiki/Data_Structures}} and as found in the wild in various datas sources, including but not limited to relational databases (RDB), NoSQL data stores~\cite{Cattell:SIGMOD11}, or Web data formats such as JSON, OData and RDF serialisations. 

\subsection{Tabular}
\label{sub:fds-tabular}

A tabular data shape organizes data items into a table. A table is a set of data elements (values) that are organized using a model of vertical columns (identified by their name), and horizontal rows. A table has a specified number of columns. Examples of tabular data shapes are: 
\begin{itemize}
	\item \textbf{CSV} (Comma Separated Values) files as of RFC 4180\footnote{\url{http://tools.ietf.org/html/rfc4180}}---These files are used to store tabular data, capable of storing numbers as well as text in a plain-text format that can be easily written and read by humans and software alike.
	\item \textbf{RDB} (relational databases)---A relational database is essentially a group of tables (entities). Tables are made up of columns and rows (tuples). Those tables have constraints, and relationships are defined between them. Relational databases are queried using SQL, and result sets are produced from queries that access data from one or more tables. 
\end{itemize}

\subsection{Tree}
\label{sub:fds-tree}
A tree is a non-empty set, one element of which is designated the root of the tree while the remaining elements are partitioned into non-empty sets each of which is a subtree of the root. Tree nodes have many useful properties. The depth of a node is the length of the path (or the number of edges) from the root to that node. The height of a node is the longest path from that node to its leaves. The height of a tree is the height of the root. A leaf node has no children, its only path is up to its parent.

A particular case of a tree is the \emph{key-value data shape}---a linked list of key-value pairs. Examples of tree data shapes are: 
\begin{itemize}
	\item \textbf{XML} (eXtensible Markup Language)---An open and flexible format used to exchange a wide variety of data on and off the Web. XML is a tree structure of nodes and nested nodes of information where the user defines the names of the nodes\footnote{\url{http://www.w3.org/XML/}}. 
	\item \textbf{JSON} (JavaScript Object Notation)--- A lightweight data-interchange format. It is easy for humans to read and write as well as straightforward for machines to parse and generate\footnote{\url{http://json.org/}}. 
	\item \textbf{YAML} (YAML Ain't Markup Language)---A Super-set of JSON and general-purpos data serialization language designed to be human-friendly and work well with modern programming languages for common everyday tasks\footnote{\url{http://yaml.org/}}. 
\end{itemize}

\subsection{Graph}
\label{sub:fds-graph}
A graph is a mathematical structure consisting of a set of vertexes (also called nodes), and a set of edges. An edge is a pair of vertexes. The two vertexes are called edge endpoints. A graph may be either undirected or directed. Intuitively, an undirected edge models a ``two-way'' or ``duplex'' connection between its endpoints, while a directed edge is a ``one-way'' connection, and is typically represented by an arrow.

Examples of graphs are:
\begin{itemize}
	 \item \textbf{RDF} (Resource Description Framework)--- A family of World Wide Web Consortium (W3C) specifications originally designed as a metadata model. It has come to be used as a general method for conceptual description or modeling of information that is implemented in web resources, using a variety of syntax formats\footnote{\url{http://www.w3.org/RDF/}}. 
	\item \textbf{Topic Maps}---Topic maps are an ISO standard for describing knowledge structures and associating them with information resources\footnote{\url{http://www.isotopicmaps.org/}}. 
\end{itemize}


\section{Data shapes transformations}
\label{sec:fct}
In this section we compare the possible transformations we can perform between two given data shapes. To this end, we have identified a set of features along three dimensions---the input, the output, and the transformation process---and provide motivational usage scenarios per transformation.
We acknowledge that the characterisations and the formats presented in following are neither exhaustive nor complete, however, serve as a useful starting point.

\begin{itemize}
	\item Dimension 1---concerning the \textbf{input data shape}:
	\begin{itemize}
		\item The generic data shape, e.g., tabular, tree or graph.
		\item The specific implementation of the data shape, e.g., XML, JSON, relational database, ect.
	\end{itemize}
	\item Dimension 2---concerning the \textbf{output data shape}:
	\begin{itemize}
		\item The generic data shape, e.g., tabular, tree or graph.
		\item The specific implementation of the data shape, e.g., XML, JSON, relational database, ect.
	\end{itemize}
	\item Dimension 3---concerning the \textbf{transformation process}:
	\begin{itemize}
		\item The transformation process can be \textbf{declarative} or \textbf{operational}. 
		\begin{itemize}
			\item \emph{Declarative}. There is a transformation description, the transformation is based on a language that describes the mappings between the input and output shapes.
			\item \emph{Operational}. The transformation is only based on an \emph{ad-hoc} transformation engine.
		\end{itemize}		
		\item The transformation process can have an information loss (also known as \textbf{lossy transformation}) defined by: ``all queries that are possible on the original shape are also possible on the resultant shape''. We have information loss when we change the abstraction level; this happens typically, when we transform a ``richer'' shape into a ``less rich shape'', e.g., from graph to tabular. 
	\end{itemize}
\end{itemize}

The Table~\ref{tab:shapetransformations} illustrates all possible transformations between two given data shapes and provides pointers to the respective subsections where we discuss them in further detail.

\begin{center}
\begin{table}[h!]
\label{tab:shapetransformations}
\caption{Data shapes transformations overview.}
\centering
\begin{tabular}{p{0.15\linewidth} p{0.18\linewidth} p{0.18\linewidth} p{0.18\linewidth}}
\hline
\emph{from/to} & \textbf{tree} & \textbf{tabular} & \textbf{graph}\\
\hline
\textbf{tree} & cf. Section~\ref{sub:tds-tree-tree} & cf. Section~\ref{sub:tds-tree-tabular} & cf. Section~\ref{sub:tds-tree-graph}\\
\hline
\textbf{tabular} & cf. Section~\ref{sub:tds-tabular-tree} & cf. Section~\ref{sub:tds-tabular-tabular} & cf. Section~\ref{sub:tds-tabular-graph}\\
\hline
\textbf{graph} & cf. Section~\ref{sub:tds-graph-tree} & cf. Section~\ref{sub:tds-graph-tabular} & cf. Section~\ref{sub:tds-graph-graph}\\
\hline
\end{tabular}
\end{table}
\end{center}

\subsection{Tree--Tree}
\label{sub:tds-tree-tree}
In this case, the transformation takes as input a given tree and outputs another tree. Let us suppose we have a set of XML documents that contain the description of the transactions of a company, and we need to submit these tin JSON files instead of XML, so we need to perform  a transformation from tree to tree. Examples of these transformations are:
\begin{itemize}
	\item XML to XML. An XSLT that turns a DocBook\footnote{\url{http://www.docbook.org/}} file into XHTML. 
	\item XML to JSON. A program that turns a XML file into JSON, or, for example via XSLT. 
\end{itemize}





\subsection{Tree--Tabular}
\label{sub:tds-tree-tabular}
This transformation takes as input a tree and outputs a tabular. Let us suppose we have a set of XML that contain the description of the transactions of a company, and we need to submit these to an entity such as a government agency that works with CSV files instead of XML, so we need to perform  a transformation from tree to tabular. Examples of these transformations are:
\begin{itemize}
	\item XML to RDB:
		\begin{itemize}
			\item In~\cite{Flik:GVSU09} a technique is described to transform XML into a RDB. The thecnique relies on the XSD of the XML. 
			\item The connect xml-2-db tool\footnote{\url{http://www.skyhawksystems.com/users_guide/runningxml2db.htm}} relies on mapping files.
		\end{itemize}
	\item XML to CSV: 
		\begin{itemize}
			\item XSLT\footnote{\url{http://www.w3.org/TR/xslt}}.
			\item Scripts\footnote{For example, \url{http://www.ricebridge.com/xml-csv-convert.htm}}.
		\end{itemize}
\end{itemize}

\subsection{Tree--Graph}
\label{sub:tds-tree-graph}
This transformation takes as input a tree and outputs a graph. Let us suppose we have a set of XML document that contain the description of the transactions of a company, and we need to submit these to an entity such as a government agency that works with RDF for integration purposes.In this setup, we need to transform from tree to graph. Examples of these transformations are:
\begin{itemize}
	\item For example, with \emph{Gleaning Resource Descriptions from Dialects of Languages} (GRDDL)\footnote{\url{http://www.w3.org/TR/grddl/}} one can turn an OData document\footnote{\url{http://www.odata.org/}} file into a corresponding RDF representation. 
	\item Rhizomik ReDeFer\footnote{\url{http://rhizomik.net/redefer/}} that includes XSD2OWL, XML2RDF. 
	\item XSPARQL\footnote{\url{http://www.w3.org/Submission/xsparql-language-specification}} is a query language combining XQuery and SPARQL for transformations between RDF and XML.
\end{itemize}

\subsection{Tabular--Tree}
\label{sub:tds-tabular-tree}
This transformation takes as input a tabular shape and outputs a tree shape. Let us suppose we have a relational database containing transaction data of a company, and we need to submit these transactions that requires, for integration purposes, the data in XML, so we need to transform from tabular to tree. Examples of these transformations are not standardised, but there are bespoke systems such as:

\begin{itemize}
	\item XML representation of a relational database\footnote{\url{http://www.w3.org/XML/RDB.html}}.
	\item XMLSpy Relational Database Integration\footnote{\url{http://www.altova.com/xmlspy/database-xml.html}}.
	\item CSV-to-XML\footnote{\url{http://csv2xml.sourceforge.net/}}.
\end{itemize}

\subsection{Tabular--Tabular}
\label{sub:tds-tabular-tabular}
This transformation takes as input a tabular and outputs a tabular. Let us suppose we have a relational database that contain the description of the transactions in our company. We need to display these transactions in the company web page. To this end, we have to transform from a tabular (RDB) to a tabular (web page). Examples of these transformations are:

\begin{itemize}
	\item RDB to RDB: SQL SELECT. 
	\item CSV to RDB: relying on a particular DBMS import tool.
\end{itemize}

\subsection{Tabular--Graph}
\label{sub:tds-tabular-graph}
This transformation takes as input a tabular and outputs a graph. Let us suppose we have a relational database that contain the description of the transactions in our company, and we need to submit these transactions into the central office in London. For integration purposes the central office is using RDF, so we need to transform from tabular to graph. Examples of these transformations are:

\begin{itemize}
	\item RDB to RDF: 
		\begin{itemize}
			\item W3C's RDB2RDF activity\footnote{\url{http://www.w3.org/2001/sw/rdb2rdf/}}: Direct Mapping and R2RML, a language for expressing customized mappings from relational databases to RDF datasets.
		\end{itemize}
	\item CSV to RDF:
		\begin{itemize}
			\item XLWrap - language
			\item TopBraid - tool
			\item RDF extension of Google Refine - tool - GUI
		\end{itemize}
	\item RDB to Topic maps \cite{Neidhart_2009}.
		
\end{itemize}

\subsection{Graph--Tree}
\label{sub:tds-graph-tree}
This transformation takes as input a graph and outputs a tree. Let us suppose we have an RDF dataset for representing the statistical information of a company and we need to transfer this information to an XML-based format such as PC-Axis, used by the Irish CSO. Examples of these transformations are:

\begin{itemize}
	\item Turning RDF to XML\footnote{\url{http://www.w3.org/wiki/ConverterFromRdf}}.
	\item XSPARQL\footnote{\url{http://www.w3.org/Submission/xsparql-language-specification}} is a query language combining XQuery and SPARQL for transformations between RDF and XML.
	\item Geo2KML\footnote{\url{http://graphite.ecs.soton.ac.uk/geo2kml/}} web service which converts RDF to KML suitable for showing on Google Earth \& Maps.
\end{itemize}

\subsection{Graph--Tabular}
\label{sub:tds-graph-tabular}
The transformation takes as input a graph and outputs a tabular. Let us suppose we have an RDF dataset for representing the statistical information of a company and want to use Google Charts for visualising it. This requires a tabular representation in CSV and therefore we have to perform a transformation from graph to tabular. Examples of these transformations are:

\begin{itemize}
	\item SPARQL SELECT
	\item \emph{ad-hoc} conversion scripts.
\end{itemize}

\subsection{Graph--Graph}
\label{sub:tds-graph-graph}
This transformation takes as input a graph and outputs a graph. Let us suppose we have an RDF dataset for representing the statistical information of a company, expressed in the W3C RDF Data Cube vocabulary\footnote{\url{http://www.w3.org/TR/vocab-data-cube/}}. Now, further assume that someone is still using the deprecated SCOVO\footnote{\url{http://purl.org/NET/scovo#}} vocabulary for representing the statistical information. Therefore we need to transform our data expressed in RDF Data Cube to SCOVO. In this case, we have to perform a transformation from a graph to graph. Examples of these transformations are:

\begin{itemize}
	\item RDF to RDF: 
		\begin{itemize}
			\item SPARQL CONSTRUCT.
			\item R2R\footnote{\url{http://www4.wiwiss.fu-berlin.de/bizer/r2r/}}.
		\end{itemize}
	\item JSON to RDF: JSON to RDF web service\footnote{\url{http://graphite.ecs.soton.ac.uk/rdf2json/}}
	\item RTM\footnote{\url{http://www.ontopia.net/topicmaps/materials/rdf2tm.html}} is a vocabulary that can be used to describe the mapping of an RDF vocabulary to topic maps in such a way that RDF data using that vocabulary can be converted automatically to topic maps.
\end{itemize}

\subsection{Summary}
\label{sub:tds-summary}
In Table~\ref{tab:comparativeFramework} we provide a summary of the data shapes transformation and their characteristics.

\begin{table}[ht!]\small
\caption{Data shapes transformations comparison.}
\label{tab:comparativeFramework}
\centering
\begin{tabular}{|p{0.20\linewidth}|p{.20\linewidth}|p{.16\linewidth}|p{.12\linewidth}|p{.34\linewidth}|}
\hline
\rowcolor[gray]{.8}
\ Input & Output & Nature & Lossy?  & Standard \\
\hline
Tabular (RDB) & Tabular (RDB) & Declarative & No  & SQL \\
\hline
Tabular (RDB) & Tree (XML) & Operational  & No  & No \\
\hline
Tabular (RDB) & Graph (RDF) & Declarative & No  & RDB2RDF \\
\hline
Tree (XML) & Tabular (RDB) & Operational & No  & No \\
\hline
Tree (XML) & Tree (XML) & Declarative & No  & XSLT \\
\hline
Tree (XML) & Tree (XML) & Declarative  & No  & XSLT \\
\hline
Graph (RDF) & Tabular (RDB) & Declarative  & Yes  & SPARQL SELECT \\
\hline
Graph (RDF) & Tree (XML) & Declarative & Yes  & No \\
\hline
Graph (RDF) & Graph (RDF) & Declarative & No  & SPARQL CONSTRUCT \\
\hline
\end{tabular}
\end{table}

\normalsize


\section{Discussion}
\label{sec:rac}
Motivated by our experiences gathered in data integration projects as well as in standardisation activities within W3C we wanted to provide a principled overview on the fundamental data shapes and transformations between them. 
Summarising, we can state the following:

\begin{itemize}
	\item We can perform (loss-less) data shape transformations between certain shapes.
	\item A number of data shape transformations are already standards or in the process of being standardised, including:
	\begin{itemize}
		\item For RDB2RDF, see R2RML and Direct Mapping.
		\item For XML2XML, see XSLT.
		\item For XML2RDF, see GRDDL.
	\end{itemize}
	\item We found that some data shape transformations are declarative in nature and it would be interesting to learn if others can and should be expressed declaratively as well.
	\item To this end, we have not taken provenance information in the transformation process into account. Again, this is something worthwhile to follow up on.
	\item In certain cases we have to deal with lossy transformations. A more systematic study of these cases, including an assessment of the implications concerning the data integration process is subject to future research.
\end{itemize}

We hope that the report in the current form is useful for both researchers and practitioners alike and consider it as one contribution in helping to establish a discussion around data shapes and their transformations in order to advance the state of the art.

\section{Acknowledgements}
The authors are grateful for the support received through the European Commission FP7 ICT projects BIG--\emph{Big Data Public Private Forum} (Grant Agreement No. 257943) as well as LATC--\emph{LOD Around-The-Clock} (Grant Agreement No. 256975).

\bibliographystyle{alpha}
\bibliography{datashapes}

\end{document}